\documentstyle[12pt]{article}

\begin{document}

%\widetext

%\draft

\title{\Large \bf Solitary vortex couples in viscoelastic Couette flow}
\author{Alexander Groisman and Victor Steinberg\\
\\
\it Department of Physics of Complex Systems,\\
\it Weizmann Institute of Science,\\ 
\it Rehovot, 72100, Israel}
\maketitle
\vskip 1cm
 
\begin{abstract}
We report experimental discovery of a localized structure, which is of a new type for dissipative systems.
It appears as a solitary vortex couple ("diwhirl") in Couette flow with
highly elastic polymer solutions. A unique property of the diwhirls is that they are {\it stationary},
in contrast to the usual localized wave structures in both Hamiltonian and dissipative systems
which are stabilized by wave dispersion. 
It is also a novel object in fluid dynamics - a couple of vortices that build a single entity somewhat 
similar to a magnetic dipole. The diwhirls arise as a result of a purely elastic instability through a hysteretic transition at 
negligible Reynolds numbers. It is suggested that the vortex flow is driven by the same forces that cause the 
Weissenberg effect. The diwhirls have a striking asymmetry between the inflow and outflow, which is also
an essential feature of the suggested elastic instability mechanism.   
\end{abstract}

\newpage

Stable spatially localized structures have been observed in many conservative and weakly
 dissipative systems(1,2). They have a form of waves with
a spatially modulated amplitude. These solitary waves are stabilized by a balance between the wave 
dispersion and nonlinearity. In weakly dissipative systems
they usually arise as a result of a hysteretic transition, while the dissipation is 
selecting a unique amplitude profile and group velocity(3). 
Quite recently, oscillatory solitary structures have been found in strongly dissipative parametrically 
driven systems(4,5). We report here the discovery of a new type of localized structure which 
is {\it stationary} and appears as a
couple of vortices in rotating Couette flow. These solitary vortex couples arise as a result of a purely 
elastic instability (at very low Reynolds numbers), if 
the working fluid is a highly elastic polymer solution. Like in Ref.4,5, the system is highly dissipative
and the transition is strongly hysteretic.

A Couette-Taylor (CT) column is a simple arrangement of two coaxial cylinders with a working
fluid in the annular gap between them. If the fluid is Newtonian, the outer cylinder is stationary, and the inner 
cylinder is rotating, at some rotation velocity $\Omega_T$ a pattern of toroidal vortices
appears on the background of the basic purely azimuthal flow (Couette flow)(6). These Taylor vortices are 
stationary and build an axially periodic axisymmetric array. They arise, because the balance between
the centrifugal force experienced by the rotating fluid, and the radial pressure gradient is unstable
with respect to radial motion of the fluid. This instability is locally symmetric with respect to
the fluid motion outwards (outflow) and inwards (inflow), so inflow and outflow in the Taylor 
vortices look rather similar.

The behavior of viscoelastic liquids in the CT geometry can be quite different from that of the usual
Newtonian fluids. If, for example, a vertical rotating rod is inserted in a beaker with a highly elastic 
polymer solution, the liquid starts to climb up on it, instead of being pushed outward by the centrifugal force. 
The reason of this "rod climbing" (Weissenberg effect) (7,8)
is, that the rod rotation produces a shear flow, which stretches the polymer molecules around the rod 
in the azimuthal direction. These elongated molecules act as stretched rubber rings that push the 
liquid towards the rod ("hoop stress"). In other words, one can say that stretching of the polymer molecules 
along the streamlines leads to a negative normal stress difference 
$N_1=\sigma_{\theta\theta}-\sigma_{rr}$, where $r$, $\theta$ and $z$ are cylindrical coordinates. Since the cylindrical 
geometry is curvilinear, this negative $N_1$ produces a volume force acting inwards in the radial direction that causes
the rod climbing.

Therefore, it is natural to suppose that for a highly elastic polymer solution the character of instability in the Couette
flow will be different
 as well. In particular, shear rate and solution elasticity, instead of the fluid velocity 
and density, should now come into play. A possible instability mechanism was proposed by Larson, Shaqfeh and Muller(9). 
To describe the polymer solution rheology, they used the elastic dumbbell model(7), where a 
polymer molecule is modeled  by two beads connected by a spring. For the Couette flow
it gives $N_1\sim\langle R_r^2\rangle(\tau\dot{\gamma}_{r\theta})^2$. Here $\langle R_r^2\rangle$ is the average 
square of the $r$-component of the vector $\vec R$ connecting the beads of a dumbbell,$\dot{\gamma}_{r\theta}$ 
is shear rate and 
$\tau$ is the polymer relaxation time. $\langle R_r^2\rangle$ is proportional to the temperature of the liquid 
and is not affected by the Couette flow. The non-dimensional combination $\tau\dot{\gamma}_{r\theta}$ 
is called the Deborah number $De$, so, the azimuthal stretching of the dumbbells and the hoop stress are proportional
to $\langle R_r^2\rangle$  and $De^2$. Any radial fluid motion in the CT column implies regions 
with positive ${\partial v_r \over \partial r}$ that corresponds to an elongational flow. Such a flow 
stretches the dumbbells in the radial direction and increases $\langle R_r^2\rangle$. This radial stretching 
is coupled to the strong primary shear flow and causes additional azimuthal elongation and growth of $N_1$ 
and the hoop stress. This increased hoop stress reacts back on the flow driving the radial motion.

An important feature of this mechanism that was not discussed in Ref.9, is its asymmetry
with respect to the radial motion outwards and inwards.
 A fluid particle that starts its radial motion in any direction should
be first accelerated. This implies positive ${\partial v_r \over \partial r}$, radial stretching of the dumbbells,
and local growth of $N_1$ and hoop stress. This increased local hoop stress will accelerate a fluid particle moving
inwards and slow down the outward motion. Therefore, one can expect the vortex patterns to have major
differences between the inflow and the outflow. 

We conducted our experiments in a temperature-controlled CT column with the inner cylinder radius $R_1=34$mm, the gap 
$d=7$mm, and the length $L=516$mm.
As an elastic liquid, we used a 300ppm solution of high molecular weight PAAm(9) in a viscous Newtonian solvent which
was a 63\% solution of saccharose in water. Solution viscosity and relaxation time were measured with the aid of
a commercial viscometer(10). 
In the explored temperature region of $5-37.5^{\circ}$C the ratio of the solution viscosity 
to the solvent viscosity was practically constant at $\eta/\eta_s=1.82$, while $\eta_s$ changed from 0.35 to 2.9 Ps.
The polymer relaxation time $\tau$
followed the theoretically predicted $\tau\sim\eta_s/T$ ($T$ is the absolute temperature). Thus, by changing the
temperature in CT column from 37.5 to $5^{\circ}$C we could change $\tau$ from 0.35 to 3.3s.

The sequence of flow patterns in the CT column was the same in the whole studied region of $\tau$ (Fig.1).
As the rotation velocity was raised, at some critical value $\Omega_0$ the basic Couette flow became unstable
and a pattern of chaotically oscillating vortices appeared in the column (Fig.1A). The transition 
was abrupt and strongly hysteretic. If $\Omega$ were lowered afterwards, the flow pattern evolved and a few
other types of patterns (Fig.1B-E) were observed until the Couette flow was finally recovered at a rotation velocity 
$\Omega_c$ that could be as low as $0.4\Omega_0$. The subject of this Letter is the pattern shown in Fig.1E that
appears as a collection of stationary localized structures, which we call solitary vortex couples  or "diwhirls".  

A typical pattern of diwhirls in the CT column is shown in Fig.2.  Fig.3A presents a typical dependence of the 
radial velocity $v_r$ on the axial position. Streamlines of a diwhirl in the $rz$-plane shown in Fig.3B 
somewhat resemble the field lines of a magnetic dipole. One can see that every diwhirl is really a couple of vortices
having a common core - a narrow region, about $d/2$ in width, of fast fluid motion {\it inwards}. The outflow 
is slow and spreads over regions of about $2.25d$ at the both sides of the core, decaying at the vortex 
edges. The diwhirls are, thus, localized within regions of about $5d$ along the column axis, the flow between 
them being just the unperturbed Couette flow. The velocity profiles of different diwhirls are strikingly similar.
They are symmetric (which implies that the vortices in the diwhirls are just mirror images of each other), 
have the same width and height, and even the same peculiarities in the outward velocity -- local minima at
about $0.75d$ from the center. This form of the diwhirl velocity profile was also independent of $\tau$. 

The axial position of an isolated diwhirl can be quite stationary, changing  by less than 0.1mm per hour. 
If, however, the distance between two diwhirls is less than about $5d$, they move towards each other and finally
coalesce (Fig.4). The motion of the close diwhirls towards each other indicates that overlapping of vortices belonging 
to different couples leads to the diwhirl attraction. The final diwhirl separation depends on the flow history. 
If the rotation velocity is quenched from above $\Omega_0$
to slightly above $\Omega_c$, at first a lot of closely spaced diwhirls are produced. The diwhirls then start to move
towards each other and coalesce. This merging continues until the distance between neighboring diwhirls reaches the
"safe" value of about $5d$. In general, the number of diwhirls at fixed $\Omega$ varied from 
one to about a dozen depending on the flow history.

Strong evidence for the elastic origin of diwhirls is that in the explored region of $\tau$,
the Deborah number at $\Omega_c$, $De_c=\Omega_c \tau R_1/d$,  remained constant 
(up to 5\%) at a value of about 11. It means that the diwhirls always decayed at the same value of
the hoop stresses. In the same region of $\tau$, the Reynolds number at $\Omega_c$ decreased 
from 33\% to 0.4\% of its critical value corresponding to $\Omega_T$, making the inertial instability 
mechanism completely irrelevant. Further, the maximal radial velocity in diwhirls
was found to be inversely proportional to the elastic relaxation time, so 
that $v_{r, max}\simeq 0.5d/\tau$ at $\Omega_c$.

The major asymmetry between the inflow and outflow in diwhirls (Fig.3) was conceived above from the general 
properties of the elastic instability mechanism. The forces driving the diwhirl flow can be understood in more detail
from the following arguments.
Although in the laboratory frame the flow in a diwhirl appears as stationary, in the reference frame of moving fluid
(Lagrangian coordinates)  the rate of strain changes periodically as a fluid particle moves along the 
flow lines (Fig.3B). Since conformation of a polymer molecule depends on history of deformations of the
fluid element inside which the molecule is situated, it is the Lagrangian coordinates that should be used for
estimation of the elastic stresses. When a fluid particle starts its 
radial motion, it is in a region of positive ${\partial v_r \over \partial r}$ in both inflow and outflow (Fig.3B).
$v_r$ becomes maximal near the middle of the gap, and after crossing the point of maximal $v_r$
the fluid particle enters the region of negative ${\partial v_r \over \partial r}$ and contractional flow.
The characteristic time of these variations in ${\partial v_r \over \partial r}$ experienced by the particle is just $d/v_{r}$, 
where $v_{r}$ is a typical radial velocity. In the diwhirl outflow the radial motion is
slow, so that $d/v_{r}\gg \tau$ and the radial elongation of the polymers
always corresponds to the current ${\partial v_r \over \partial r}$. It implies that the average elongation
across the gap is zero, since the regions of elongational flow (positive ${\partial v_r \over \partial r}$)
are exactly compensated by the contractional flow regions. Therefore, the additional hoop stresses produced 
by the outflow are averaged to zero, when
integrated across the gap, and have small influence on this slow flow. If, however, the radial flow is fast 
enough, so that $d/v_{r}\simeq \tau$, like in the diwhirl inflow, there exists a significant phase lag between 
${\partial v_r \over \partial r}$ and the radial polymer elongation. Then $\langle R_r^2\rangle$ depends not 
only on ${\partial v_r \over \partial r}$ but also on its time integral. The latter is always positive, 
since the elongation always comes before the contraction as a fluid particle moves along a radius. Thus, the additional
hoop stress averaged across the gap is positive in this case, which results in a radial force that acts
in the inward direction and drives the inflow. 

The narrow diwhirl core turns out to be the region where the energy is pumped into diwhirls. In the Lagrangian coordinates 
$v_r/d$ plays a role of frequency, which should be large 
enough to assure a non-zero average radial elongation. That is why diwhirls arise as a result of a hysteretic transition and 
have finite $v_r$ before their decay at $\Omega_c$. Analyzing statistical distributions of $v_r$ in the chaotic 
oscillatory flows shown in Fig.1A-D, we found that the major asymmetry between the inflow and the outflow is present 
there as well. Therefore, we believe that this asymmetry is a general feature of the flow instabilities driven by the
hoop stress and the proposed instability mechanism has wide applicability.
 
Pattern formation in many dissipative systems has been successfully described by the amplitude equation(1).
This equation, however, does not have stationary localized solutions and, thus, cannot be adequate 
for the diwhirls. Nevertheless, such solutions 
can exist if the amplitude equation describes a hysteretic transition and is coupled to another dynamic equation for a 
slow mode(12). In our case, this slow mode could represent the elastic stresses which drive the fluid motion.

\newpage
\begin{center} {\bf Figure captions.} \end {center}
\begin{enumerate}

\item{Various flow patterns that are observed as $\Omega$ decreases from $\Omega_0$ to $\Omega_c$.
The flow profile across the gap in $r-z$ cross-section is shown. The top and the bottom of each strip correspond
to the outer and the inner cylinder, respectively. To visualize the flow, we used a very small amount of light reflecting
flakes (0.1\% of Kalliroscope fluid). A laser beam expanded to a sheet of light parallel to the column axis 
 was used for illumination. ({\bf A}) $\Omega=\Omega_0$. Chaotically oscillating vortex motion in the whole gap. This
pattern ("Disordered oscillations") was described in Ref.10. ({\bf B}) $\Omega=0.75\Omega_0$. Vortices are
mostly near the inner cylinder except for the regions near black spindle-shaped cores. ({\bf C}) $\Omega=0.69\Omega_0$. The 
pattern become spatially inhomogeneous. The oscillating vortices are localized inside separate strips with a core in the 
middle. ({\bf D}) $\Omega=0.59\Omega_0$. The oscillatory strips become narrower. ({\bf E}) $\Omega=0.48\Omega_0$. 
Stationary vortex structures (which abruptly decay at $\Omega_c=0.47\Omega_0$).}

\item{A photograph of the CT column with a diwhirl pattern. The flow was visualized by addition of a small amount of
light reflecting flakes (0.6\% of Kalliroscope) in the ambient illumination. Diwhirls appear as randomly spaced axisymmetric
dark rings. The dark color here, just as the dark color of the spindle-shaped cores of the diwhirls in Fig.1, indicates 
regions of intensive radial flow. A similar pattern was reported in a different polymer solution(12).}

\item{({\bf A}) Radial component of the fluid velocity $v_r$, measured at constant radius (near the middle of the
gap, where $v_r$ is maximal), as a function of position along the column axis. The velocity was measured by a laser 
Doppler velocimeter (LDV). ({\bf B}) Schematic drawing of flow lines in a diwhirl as it follows from the LDV measurements
and Fig.1E.}

\item{The consecutive stages of coalescence of two closely spaced  diwhirls (the visualization technique was the same as in
Fig.1). The energy of vortices that disappear (one vortex from each couple) is first transferred to a wavy
motion ({\bf D}) and than dissipates ({\bf E}). It is quite noticeable that the "daughter" diwhirl has the
same shape as its both "parents", which is another manifestation of the universality of the diwhirl profiles. One can also
see that the daughter diwhirl inherited the two vortices that used to be at the outer sides of the parent diwhirls, while
those two which used to be at their inner sides just annihilated.}\\
\end{enumerate}
\newpage
\centerline{\bf References}
\begin{enumerate}

\item{M.C.Cross and P.C.Hohenberg, {\it Rev. Mod. Phys.} {\bf 65 }, 851 (1993).}

\item{A.C.Newell, {\it Solitons in Mathematics and Physics }, 
(Society for Industrial and Applied Mathematics, Philadelphia, 1985).}

\item{S.Fauve and O.Thual, {\it Phys. Rev. Lett.} {\bf 64}, 282, (1990).}

\item{O.Lioubashevski, H.Arbel, J.Fineberg, {\it Phys. Rev. Lett.} {\bf 76}, 3959 (1996).}

\item{P.B.Umbanhowar, F.Melo, H.L.Swinney, {\it Nature } {\bf 382}, 793 (1996).}

\item{G.I.Taylor, {\it Philos. Trans. R. Soc. London.} {\bf A223}, 289 (1923)}.
 
\item{R.B.Bird, Ch.Curtiss, R.C.Armstrong, O.Hassager, {\it Dynamics of Polymeric 
Liquids}, Vol. 1,2 (Wiley, NY, 1987).}

\item{K.Weissenberg, {\it Nature} {\bf 159}, 310 (1947)}. 
 
\item{R.G.Larson, E.S.G.Shaqfeh, S.J.Muller, {\it J. Fluid Mech.} {\bf 218}, 573 (1990).}

\item{A.Groisman and V.Steinberg, {\it Phys. Rev. Lett.} {\bf 77}, 1480, (1996).}

\item{H.Riecke, {\it Physica} D{\bf 92}, 69, (1995).}
 
\item{R.Haas and K.B\" uhler, {\it Rheol. Acta} {\bf 28}, 402, (1989).}
\end{enumerate}

ACKNOWLEGEMENTS. We thank M.Shliomis for useful suggestions. This work was supported by the Minerva Center 
for Nonlinear Physics of Complex Systems and a research grant from the Philip M. Klutznick Fund for Research.

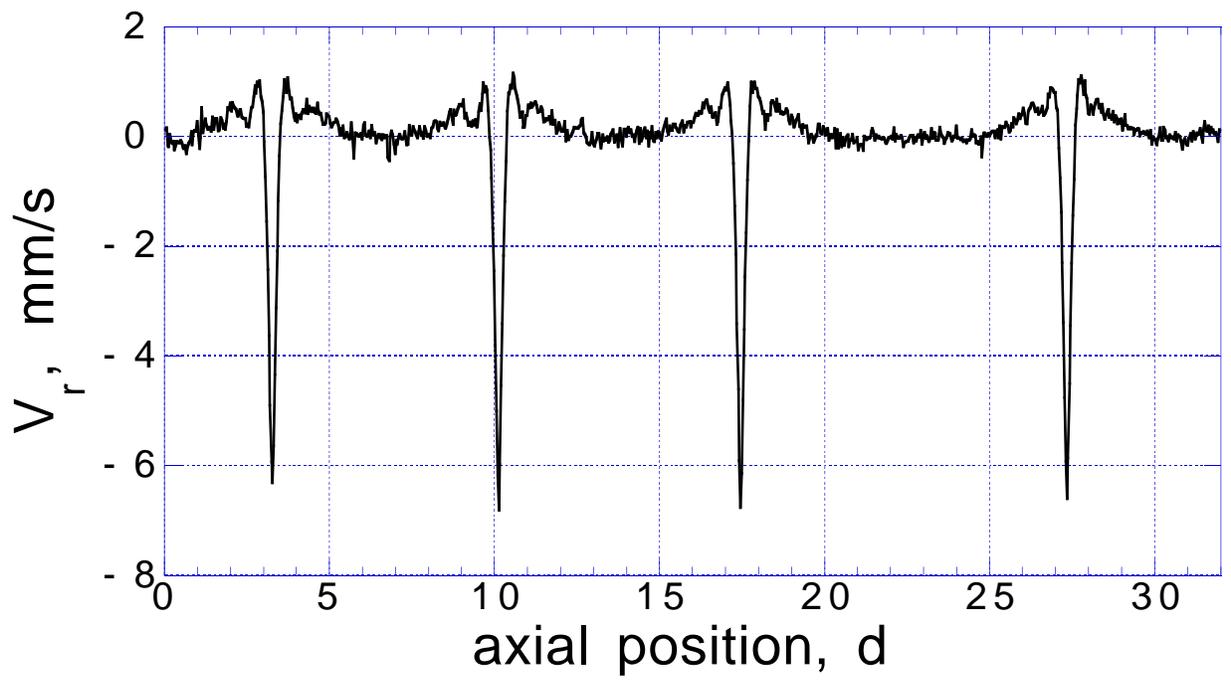
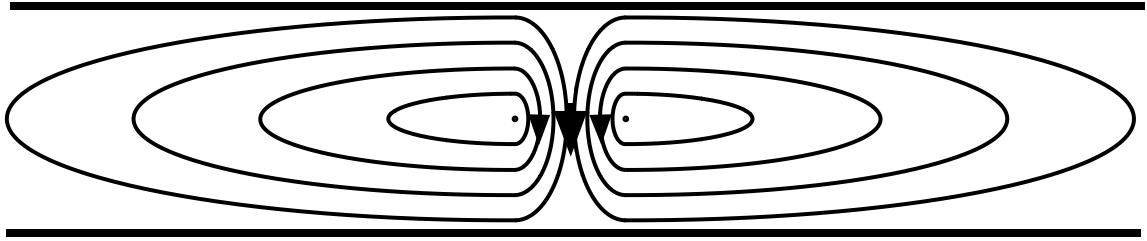
\end{document}